\documentclass[preprint,3p,twocolumn]{elsarticle}
\usepackage{graphicx}
\usepackage{mathrsfs}
\usepackage{algorithm,algorithmicx, algpseudocode}
\usepackage{amsmath}
\usepackage{caption}
\usepackage{multicol}
\usepackage[export]{adjustbox}
\usepackage{lineno,hyperref}
\usepackage{fancyvrb}

\modulolinenumbers[5]

\journal{Journal of \LaTeX\ Templates}

\let\today\relax
\makeatletter
\def\ps@pprintTitle{%
    \let\@oddhead\@empty
    \let\@evenhead\@empty
    \def\@oddfoot{\footnotesize\itshape
         {~} \hfill\today}%
    \let\@evenfoot\@oddfoot
    }
\makeatother

\begin{document}

\begin{frontmatter}

\title{Optimal routing algorithm for trips involving thousands of ev-charging stations using Kinetica-Graph$^{\vcenter{\hbox{\includegraphics[height=0.35cm]{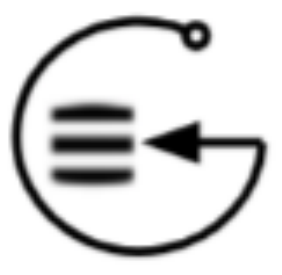}}}\dagger}$}

\author{B. Kaan Karamete\corref{cc}} 
\author{Eli Glaser\corref{}}

\cortext[cc]{\it Corresponding author: Bilge Kaan Karamete, \\ 
kkaramete@kinetica.com, karametebkaan@gmail.com\\ 
$\dagger$Kinetica-Graph: https://arxiv.org/abs/2201.02136 }
\address{Kinetica DB Inc. \break 901 North Glebe Road, Arlington, Virginia 22203}

\begin{abstract}

This paper discusses a graph based route solving algorithm to find the optimal path for an electric vehicle picking the best charging locations among thousands to minimize the total cumulative driving distance between the end points of the trip. To this end, we have devised a combinatorial optimization algorithm and a fixed storage graph topology construction for the  graph road network of the continental USA. We have also re-purposed our existing Dijkstra solver to reduce the computational cost of many shortest path solves involved in the algorithm. An adaptive and light weight spatial search structure is also devised for finding a set of prospective stations at each charging location using uniform bins and double link associations. The entire algorithm is implemented as yet another multi-threaded at-scale graph solver within the suite of Kinetica-Graph analytics, exposed as a restful API endpoint and operable within SQL. Several example trips are solved and the results are demonstrated within the context.

\noindent 

\end{abstract}


\end{frontmatter}

\section{Introduction}

\begin{figure*}
\centering
    \framebox{\includegraphics[width=\linewidth, keepaspectratio]{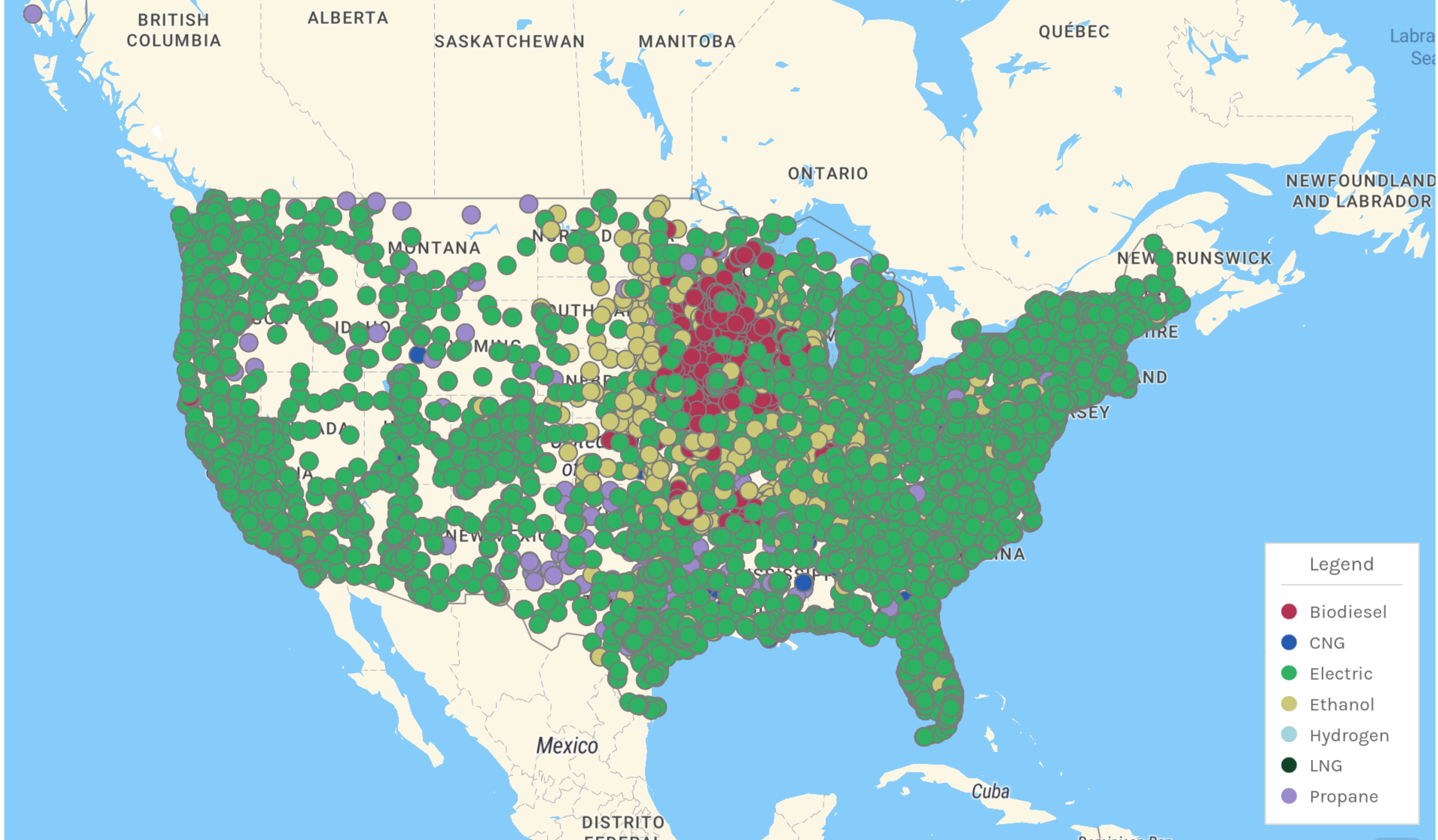}}
    \caption{Courtesy of US Dept of Energy~\cite{evstations}; alternative fuel charging stations across the US colored based on the type of the fuel shown in the legend.}
    \label{Figure:usmatchstations}
\end{figure*}

The use of the electric vehicles (EVs) increased ten fold in the last two years alone and it is estimated that the market share of EVs will increase to more than 50 percent of the passenger car market in the US by 2030~\cite{evmarket}. However, one of the key roadblocks for people to choose EVs over fossil fuel alternatives is the accessibility and the availability of the recharging stations particularly when trip durations require multiple charges due to the limited battery capacity of the EVs. In fact, there is an increased urgency in adding more recharging stations across the US, available and compatible to many brands and designs. As of 2021 there are around 45 thousand public outlets in the US~\cite{evstations} as seen in Figure~\ref{Figure:usmatchstations}. Hence, pre-planning a trip route in the best economical way possible proves to be a practical need in today's reality and a complicated challenge algorithmically considering the many factors affecting the optimal decision making process. These factors range from the sparse availability of the stations to the dynamically changing traffic conditions that have a significant impact on the energy consumption and over the result of the optimal routing between the two end points of the trip. 

We addressed this clear need by implementing a fast, practical and accurate graph based optimization solver, with parameters specific to the optimal routing problem of an EV trip involving multiple charging stops so that different capacity limits and re-charging penalties can be rolled into the optimization algorithm~\cite{msdoblog}. Various mapping and routing algorithms for EV vehicles by Mapbox, Google, TomTom, etc. are surveyed and summarized for the consumption of various EV car manufacturers, such as BMW, Tesla, Hyundai, and Nissan by Axelsson and Andreasson~\cite{evapis}. The major difference of our implementation compared to those of the referenced solvers is that our solution does not use bi-directional A-star Dijkstra between the prospective stations, and does not require finding a pivot location between charging locations. We have accomplished this by rewriting our conventional Dijkstra algorithm to fit into the SLA requirements which is critical due to the combinatorial aspect of the problem that require thousands of shortest path solves whereas many other EV routing algorithms in the literature has employed chronological-shortest path tree algorithms~\cite{liujoint,chronospt,evrouting} towards the same goal. Our optimization algorithm is summarized in Section~\ref{Section:Algorithm} and implemented using a distributed graph database hybrid with a relational DB, namely, Kinetica-Graph introduced by the authors recently~\cite{kineticagraph}.

\section{Algorithm}
\label{Section:Algorithm}

\begin{figure*}
\centering
    \framebox{\includegraphics[width=\linewidth, keepaspectratio]{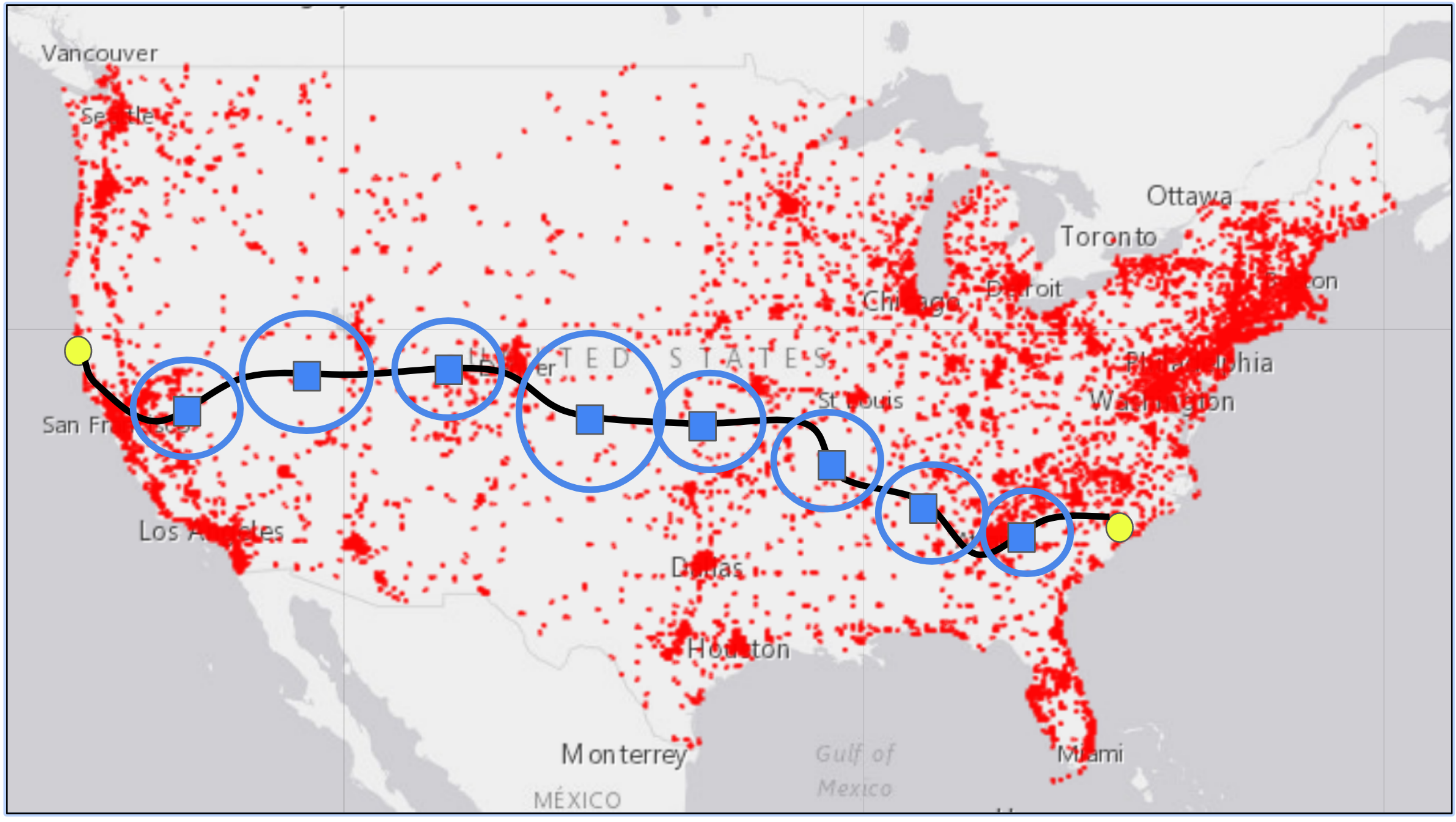}}
    \caption{First two steps of the optimization algorithm: (1) Blue rectangles are found over the shortest path route from start to target in yellow, and depicted as bases at a distance proportional to a percentage of the capacity of the EV, $Q$ by $\beta Q$, (2) Blue circles indicate the candidate stations searched and found within a threshold Dijkstra distance from their bases, by $\alpha Q$, where the coefficients satisfy teh constraint $\alpha + \beta = 1$ as stated in Equation~\ref{Equation:constraints}.}
    \label{Figure:usmatchmainalgo}
\end{figure*}

Our algorithm is based on the assumption that the most likely optimal path should be the one that is tracking the closest to the shortest path between the two end points of the trip. This is a reasonable assumption in the sense that finding nearby stations around possible stops off this path would still allow iterating over many combinations which must be among the the most ideal choices, and can be considered to be our heuristic optimization criteria. Our fall back scenario in case of the scarcity of stations around the shortest path is to increase the search radius around the stops until the desired number of candidate stations are found, which is also another parameter of our algorithm. Our experiments in running the solver over many pairs of source and target locations across the US has confirmed the effectiveness of our assumption and the algorithm, the steps of which are summarized and listed below and shown in Figure~\ref{Figure:usmatchmainalgo}.

\begin{itemize}
\item[-] Step 1. Construct a directed graph encompassing all available charging stations and road segments,
\item[-] Step 2. Run one A-star Dijsktra sssp (single source shortest path) solve to find the shortest path from source to destination,  
\item[-] Step 3. Split the path at locations where recharging is needed based on the capacity, depicted as 'bases',
\item[-] Step 4. Search for $n$ number of prospective stations (a parameter of the algorithm) around each 'base'.
\item[-] Step 5. Run shortest paths between the consecutive stations at adjacent bases.
\item[-] Step 6. Apply restrictions if the shortest path cost violates the charging capacity limit.
\item[-] Step 7. Construct a new network graph ('process' network) by adding an edge whose nodes are the consecutive prospective station pairs.
\item[-] Step 7.1. Assign the cost of the sssp solves as edge weights.
\item[-] Step 7.2. Map the sssp paths to the new edge of the network.
\item[-] Step 8. Solve one final sssp on the 'process' network from source to destination minimizing the total cost of the edges (i.e., in this case the total sum of individual trips between stations).
\item[-] Step 9. Retrieve the mapped paths of the edges in the solution path found above to concatenate with each other for the final result along with the station numbers on the output.
\end{itemize}

A directed Kinetica-Graph of the US road network is generated from OSM data~\cite{osm} using adaptive tiles and described in Section~\ref{Section:OSM}. A very lightweight spatial search structure will be demonstrated using uniform bins and a double-link-structure for associative items (stations) to each bin in Section~\ref{Section:Bins}. The special implementation of Dijkstra algorithm that is run between each pair of consecutive stations will be discussed in Section~\ref{Section:Dijkstra}. 

Forming the process network graph from pairwise shortest path runs between prospective station stops will be covered in Section~\ref{Section:Network}. The final shortest path run on this process network to pick the most optimal combination is discussed in Section~\ref{Section:Final}. Finally, a number of example routes will be shown using the solver implemented in this study along with the corresponding SQL syntax in Section~\ref{Section:Results}.

\section{Graph Creation from OSM Tiles}
\label{Section:OSM}

The graph road network of the continental US can be over 300 million edges which significantly poses a heavy computational burden on any optimization algorithm. We have formulated the generation of our Kinetica-Graph topology from the road network data available via OSM as tiles~\cite{osm,kineticagraph} by filtering out certain road types to cut down on size of the graph by half without impeding on our ability to solve between any two localities.
 
\begin{figure*}
\centering
    \framebox{\includegraphics[width=0.5\linewidth, keepaspectratio]{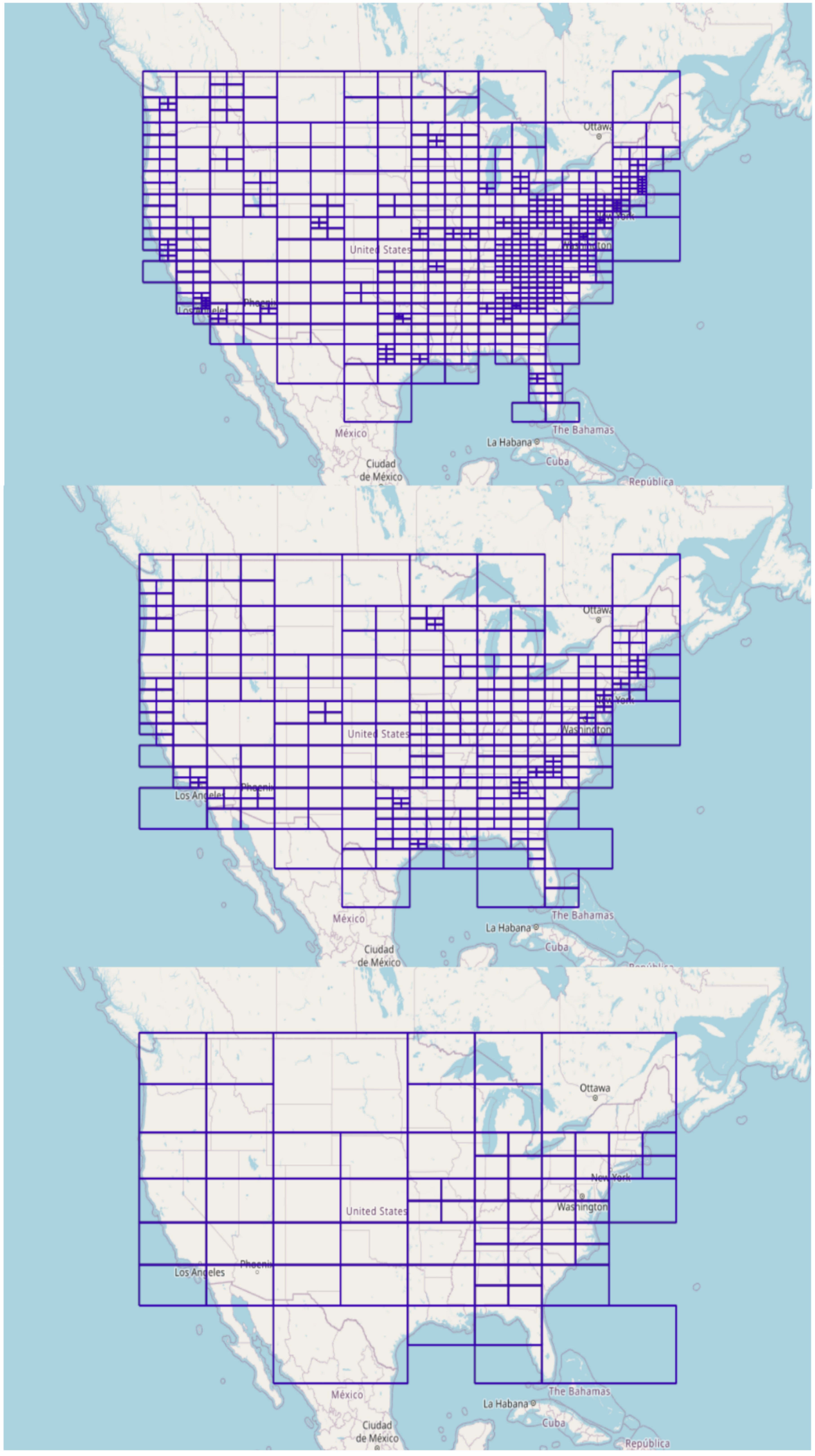}}
    \caption{Automatic partitioning of OSM road network into Kinetica-Graph as tiles; (top) 500K nodes 697 tiles, (mid) 1M nodes  371 tiles, (bottom) 2M nodes 72 tiles.}
    \label{Figure:usmatchevall}
\end{figure*}

\begin{figure}
\centering
   \includegraphics[width=\linewidth, keepaspectratio]{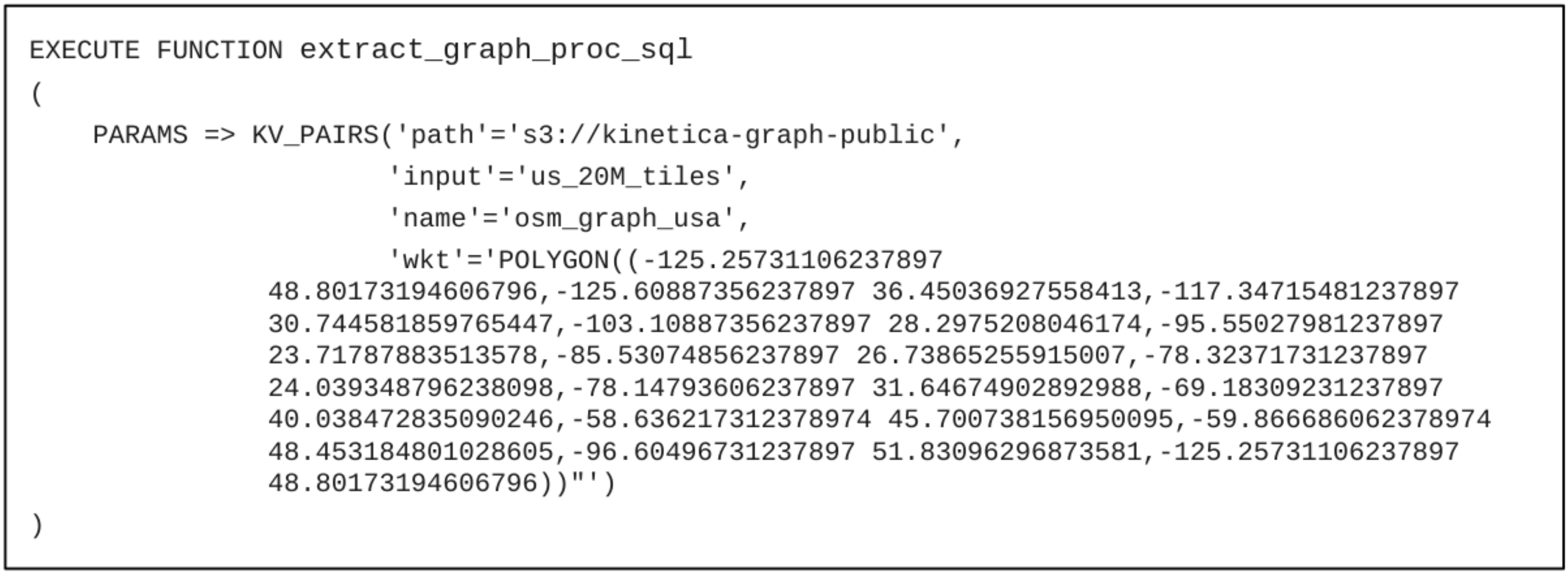}
    \caption{User defined SQL function to automatically extract OSM files and create Kinetica-Graph within the specified WKT polygon. The tiles within this threshold is embedded into the Create-Graph call depicted in Figure~\ref{Figure:usmatchcreategraph}. Tiles only share nodes and hence they can be combined without creating duplicated graph entities into a single graph shown in Figure~\ref{Figure:usmatchheatmap}.}
    \label{Figure:usmatchudf}
\end{figure}

\begin{figure*}
\centering
   \framebox{\includegraphics[width=0.4\linewidth, keepaspectratio]{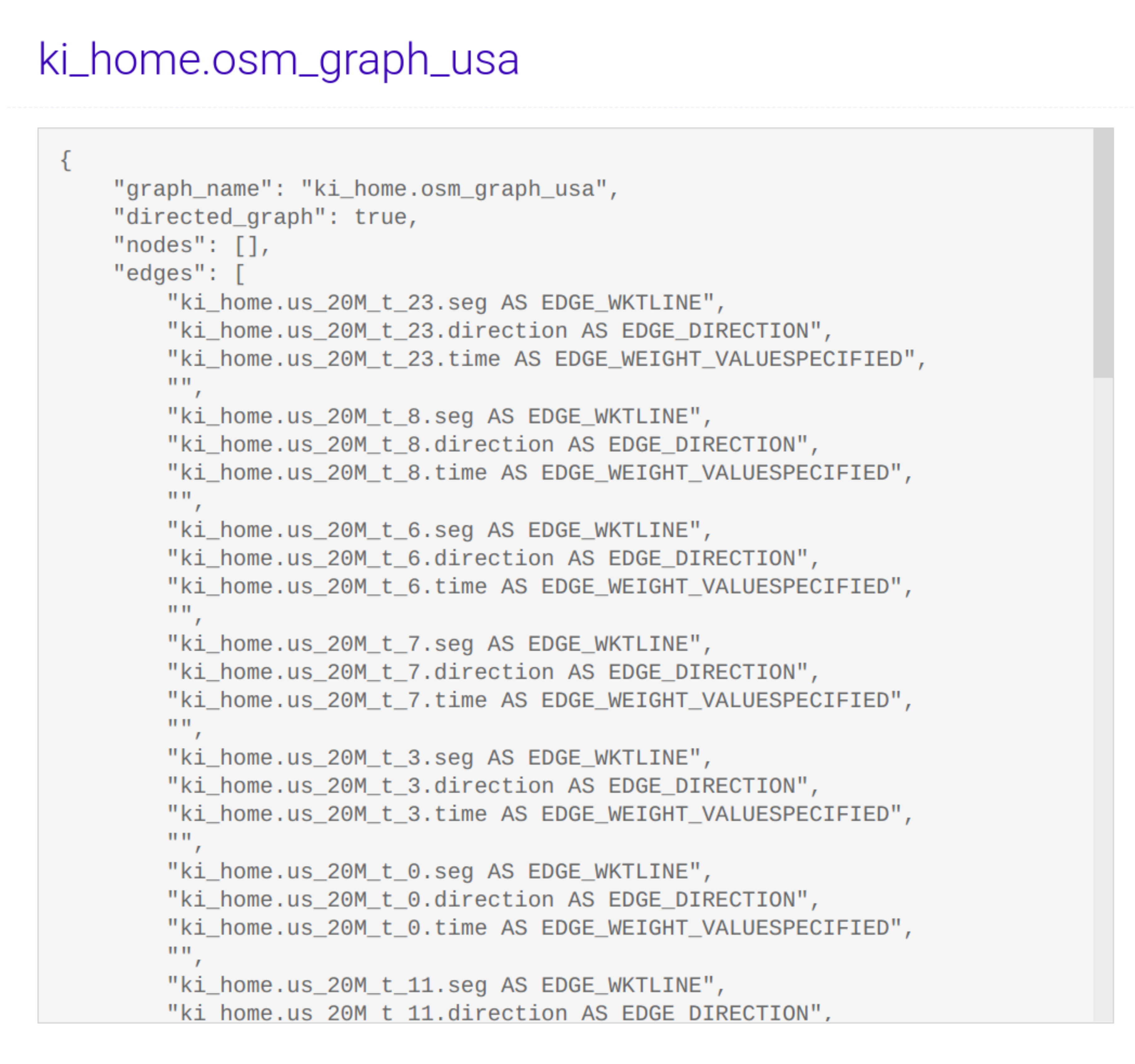}}
    \caption{Implicitly created Create-Graph request by the UDF of Figure~\ref{Figure:usmatchudf}. There are 24 tiles concatenated via the triplets of edge combinations \textit{WKTLINE, DIRECTION}, and \textit{EDGE\_WEIGHTS} coresponding to the separate DB tables with columns, \textit{seg, direction}, and \textit{time}, respectively.}
    \label{Figure:usmatchcreategraph}
\end{figure*}

\begin{figure*}
\centering  
    \includegraphics[width=0.9\linewidth, keepaspectratio]{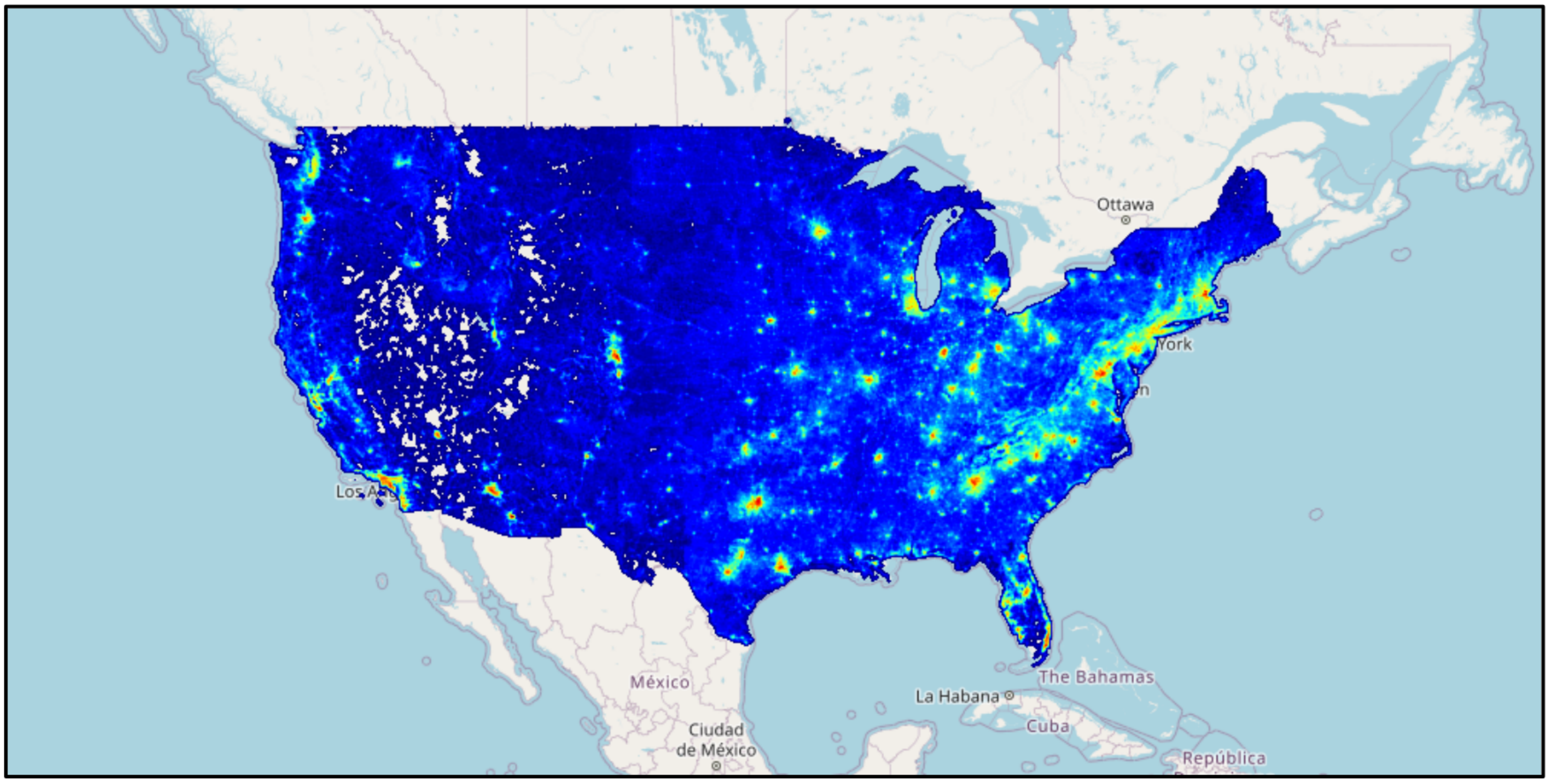}  
    \caption{The 160 million edge US road network, excluding service roads, extracted from OSM~\cite{osm}, and constructed as one directed Kinetica-Graph by combining the 24 partitioned  tiles from roughly 20 million edges in each. The graph object holds 16 GByte of memory.}
    \label{Figure:usmatchheatmap}
\end{figure*}

We have developed an automatic extraction process using Python scripts from data stored within S3 buckets, to extract OSM binary files only where the user is interested to create a Kinetica-Graph by providing an enclosing geospatial region. The other parameter is the tile threshold, in which we create a tile (a rectangular shape) input relational DB file to our create/graph as soon as the number of OSM road nodes exceeds the given threshold. Various tile division schemes can be seen in Figure~\ref{Figure:usmatchevall}. We have created an easy facility to create graphs by hiding all the complexity of the OSM network via a simple user defined SQL function (UDF) as shown in Figure~\ref{Figure:usmatchudf}. We make sure that the tiles are only connected via the duplicated nodes, with no overlapping edges. This criterion is crucial in the sense that we can then concatenate as many tiles as necessary covering the specified input bound, to create a single Kinetica-Graph object. Our Kinetica-Graph creation endpoint (Restful/C++/Python/Java/JS/R API forms available) is designed to input arbitrarily many tiles within one single call as shown in Figure~\ref{Figure:usmatchcreategraph}. This Create-Graph call request is automatically created by the UDF shown in Figure~\ref{Figure:usmatchudf}. A single graph of 160 million edges is created by combining 24 tiles, using the threshold of 20 million nodes in each. This graph requires only 16 GBytes of memory, as shown in Figure~\ref{Figure:usmatchheatmap}.

\section{Adaptive Search Bins}
\label{Section:Bins}

\begin{figure*}
\centering
    \framebox{\includegraphics[width=0.8\linewidth, keepaspectratio]{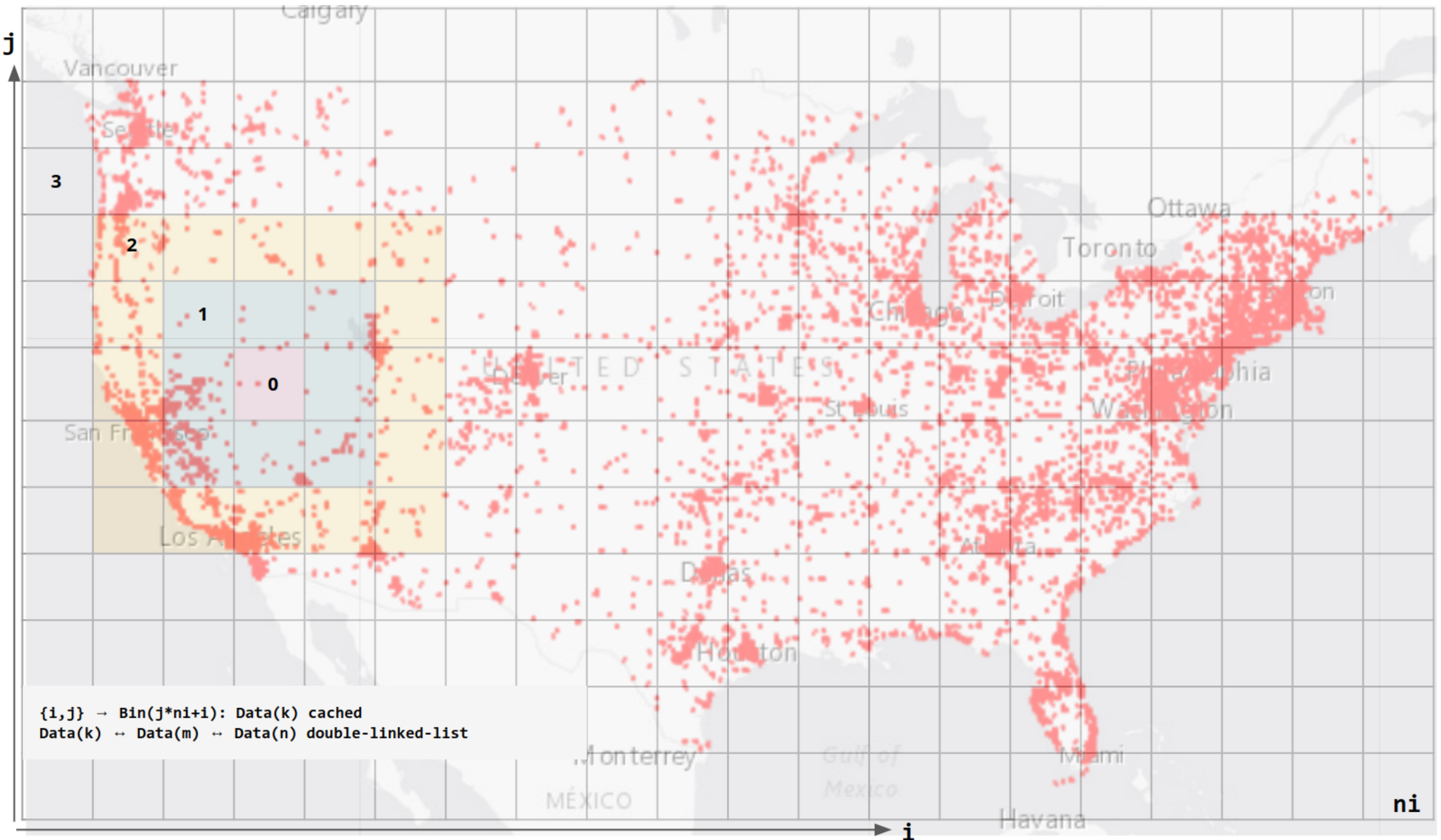}}
    \caption{Uniform bins for spatial searching: The map is divided into constant size buckets along $x$ and $y$. The number of divisions along $x$ is $ni$. Given any station coordinates ${x_i,y_i}$ we can find its ${i,j}$ indices, from which the bucket number can simply be computed by $ni*j+i$. The data, in this case the station index is cached with the bucket that it is contained. The station index is then added to the previously cached value as the next link in the linked list. This light weight of spatial indexing of the stations will only require a vector of three times the size of stations, and the map from 'containing' buckets to the cached station index.}
    \label{Figure:usmatchbins}
\end{figure*}

A uniform bin (lattice) structure is constructed with one input parameter of a delta tolerance (cell size) along x and y (longitude and latitude), respectively, defaulting to 10 kilometers. Each lattice bin is then defined by a pair of integers depicting its index on x and y, found by dividing them with the delta tolerance as shown in Figure~\ref{Figure:usmatchbins}. The bounds of the uniform bins is flexible and chosen by default to be the world coordinates (-180, to 180 along x, and -90 to +90 along y). The idea is not to use expensive adaptive structures like quad or R-trees~\cite{rtree} but a more efficient and light weight structure with the ability to grow around the search location as increasing layers (hops) when necessary during the search process. Uniform bins are also used as an associative data container; in which the only parameter used for containing association is the linearly mapped index of the bin where the data item is located. Each bin can house thousands of items, without any need for resizing and each item can only be associated with only one bin. When this one-to-many (lower order) and one-to-one (higher order) adjacency constraints are respected, only one vector of the size three times the data items is enough to spatially index the entire data as a doubly link list (dls); previous and next items, so that the removal and the addition of an item to the bin structure has constant time complexity. This lightweight structure is first devised by the author for numerical preprocessors, simulations and solvers~\cite{dls, hexdom}, and later successfully adapted for the construction of a fixed size graph topology for the Kinetica-Graph itself~\cite{kineticagraph, mapmatching}.

\begin{figure*}
\centering
    \framebox{\includegraphics[width=0.8\linewidth, keepaspectratio]{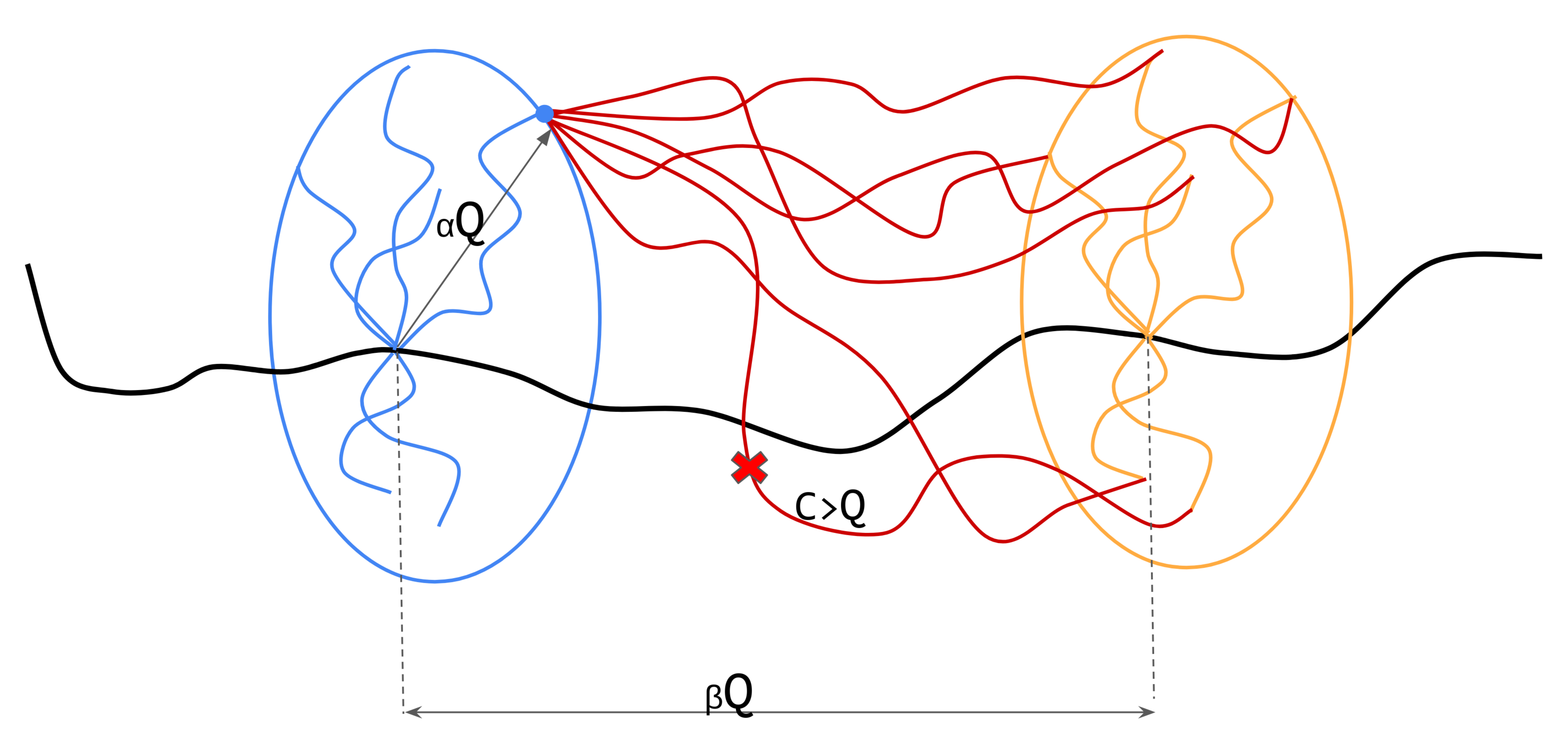}}
    \caption{Constraints on paths from one charging base to the next; All the candidate stations around each base is searched and found within $\alpha Q$ Dijkstra distance, and from each one of these prospective stations, all possible paths to the next base's prospective stations are computed. Bases are found to be within $\beta Q$ Dijkstra distance between the start and the target locations such that $\alpha + \beta = 1$. Any path whose cost is greater than the vehicle's charging capacity $Q$ is skipped. See Equation~\ref{Equation:constraints}}.
    \label{Figure:usmatchpaths}
\end{figure*}

We only need to find $n$ number of stations around each stop within a disk of $\alpha Q$, where $\alpha$ is a percentage of the charging capacity $Q$, say, $20 \%$ as seen in Figure~\ref{Figure:usmatchpaths}. At the root of each base stop location where we found by splitting the shortest path between the two end points of the trip at $\beta Q$ distances, where $\beta$ is a percentage of $Q$, say, $80 \%$, we run the Dijkstra kernel (See Equation~\ref{Equation:dijkstra_eqn}) from the base point(s) towards the adjacent stations within the disk of $\alpha Q$ as shown in blue and orange colors in Figure~\ref{Figure:usmatchpaths}, respectively. Dijkstra kernel is defined by $\mathscr{D}$ in Equation~\ref{Equation:dijkstra_eqn} as traversing a graph between two points depicted as \textit{start} and \textit{end} such that the distance that is required to reach to every node can not be greater than the sum of the distance from the traversed (incoming) node, $d_i$ and the weight of the edge $w_{ij}$ connecting the nodes. In essence, Dijkstra traversals favor the directions where the distance field at each node is the local minimum among its adjacent alternatives.

In these Dijkstra runs, we do not keep track of the traversal history as we are only interested in finding the buckets within the reachability disk (isochrone contour) within a percentage radius of the capacity $Q$. For each graph node locations within this disk, the respective grid lattice $i,j$ indices are calculated. Once the buckets are collected with one more layer around them, these buckets are used to retrieve the associated stations using the double link structure. Note that the percentages $\alpha$ and $\beta$ should sum up to be unity as the combined maximal disk between EV stations should not exceed capacity $Q$ as shown in Equation~\ref{Equation:constraints} as the constraints of our optimization algorithm.

\section{Revised Dijkstra}
\label{Section:Dijkstra}

Running shortest paths between each consecutive prospective station pairs require enormous computing resources over a directed graph of 160 million edges. We have revised our existing Dijkstra solver depicted by Equation~\ref{Equation:dijkstra_eqn} to reduce the impact of the giant graph size on the running time of the solver. Our conventional Dijkstra solver was implemented using vectors to hold the nodal distances and revisit traversal history of its priority queue implementation. However, in this specific instance, we will only need to cover within a Dijkstra disk radius of $\beta Q$, where $\beta$ is a percentage of the ev-charging capacity $Q$, i.e., we only need to run from each station the next stop's stations within this maximal disk and stop if or when we reach the target stations on the next stop from the same source at the current base stop. Hence, the use of maps in storing the nodal Dijkstra results within the disk-radius would save pre-allocating of $\approx 150$ million vector spaces every time a shortest path is to be computed. We have traced the maximal map sizes during the solve cycles to compare against the graph size, and it is found to be well within 1-5 million nodes versus 150 million, resulting in more than two orders of magnitude of memory savings. This is not be underestimated, since with such a small memory footprint per solve cycle, distributing the solves among many threads could become possible which also significantly accelerates the overall execution time of the optimization solver.

\begin{align}
d_{i} = (d_j + w_{ij})~ \mid ~w_{ij}\colon v_j \mapsto v_i,   \in N(v_{i}) \nonumber \\
\mathscr{D}_{start,end} =  \min_{v_i \in G(V,E) \mid_{start}^{end}} \Big(d_i\Big)  
\label{Equation:dijkstra_eqn}
\end{align}

\begin{align}
\mathscr{D}_{stn_i, stn_{i+1}} \in d_{max} < \beta~\cdot~Q \nonumber \\
\mathscr{D}_{base_i, stn_j} \in d_{max} <  \alpha~\cdot~Q \nonumber \\
\alpha + \beta = 1 
 \label{Equation:constraints}
\end{align}

\section{Forming the process network graph}
\label{Section:Network}

\begin{figure*}
\centering
    \framebox{\includegraphics[width=0.8\linewidth, keepaspectratio]{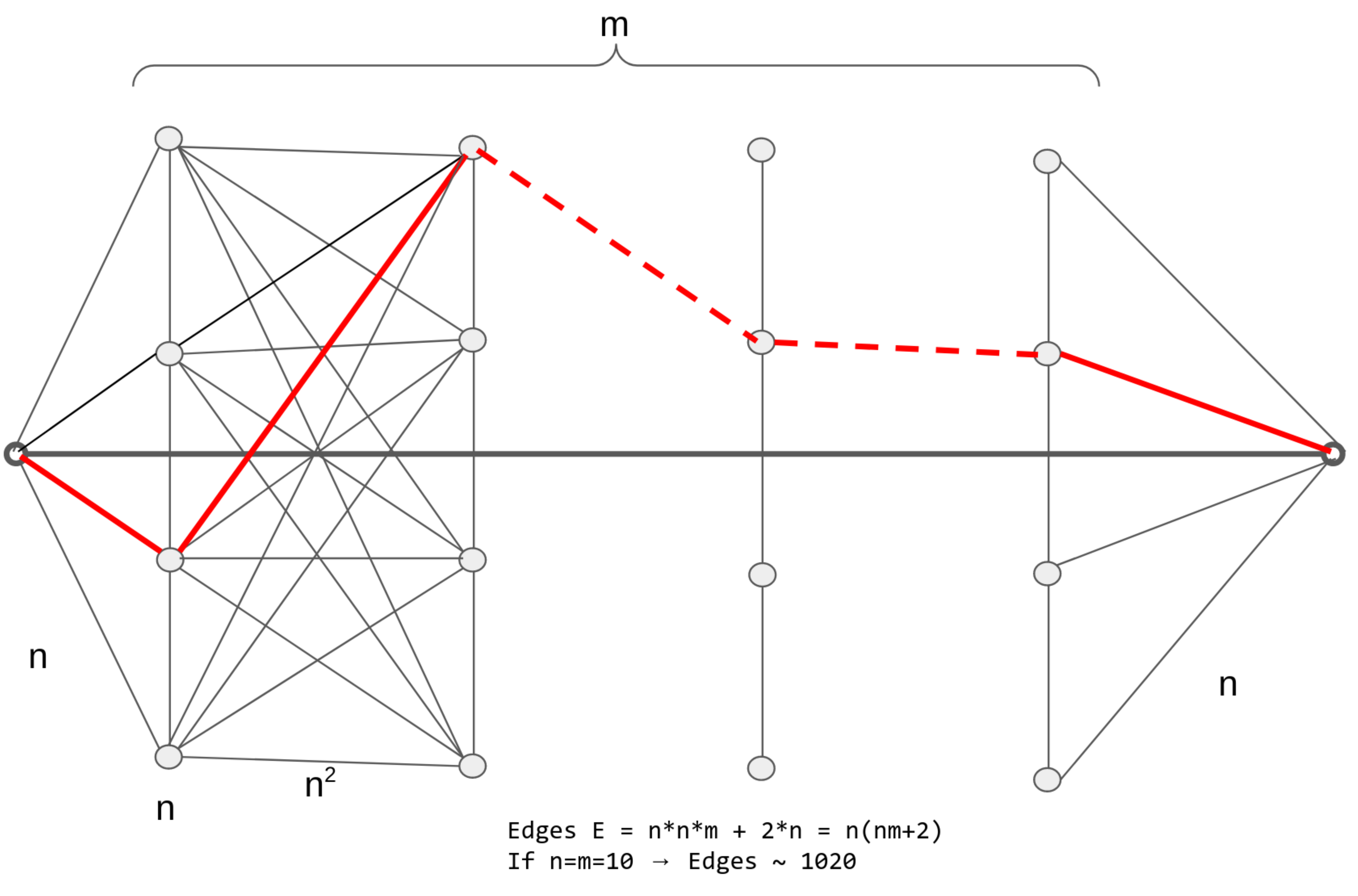}}
    \caption{The sub-network graph diagram; this process graph is generated from the shortest path runs between the prospective stations between each consecutive base pairs. Each circle depicts a prospective station along the vertical of a base vertex. A directed graph edge is depicted by a segment between two circled nodes, indicating a path between the two stations whose edge weight is computed from the Dijkstra solution's path cost. Left and right most nodes are always tied to the start and the target locations of the trip. If there are $n$ candidates per $m^{th}$ base (root location computed at a constrained distance off of the shortest path between the start and the target),  the number of graph edges in this process graph is $n(nm+2)$, i.e., for $n=m=10$, total number of edges in this graph is $E=1020$ which is also the total number of the Dijkstra runs involved in the optimization. The red line is the final result of Dijkstra run on this graph between the start and the target whose cumulative cost is the minimum. Refer to Figure~\ref{Figure:usmatchpaths} for how the paths are defined.}
    \label{Figure:usmatchnetwork}
\end{figure*}

In order to accumulate all the possible combinations from each pair of the consecutive base stations, we have devised a network sub-graph, which we call as 'process-graph' and created a directed edge between each pair of these stations as shown in Figure~\ref{Figure:usmatchnetwork}. This computational process graph diagram corresponds to the physical paths of Figure~\ref{Figure:usmatchpaths}. Between each station interval from the current base to the next base, there are a total of $n*n$ paths if $n$ is the number of the prospective stations at each base stop. There may be $m$ number of stops computed by splitting the shortest path between the two end points by the charging capacity. The overall number of process-graph edges $\mathscr{E}$ can then be computed by the following simple formula:

\begin{align}
\mathscr{E} = n(nm+2) 
\label{Equation:processedges}
\end{align}

One interesting observation of the process graph is that we have formulated all possibilities in a graph definition where the edge weights are simply the Dijkstra costs of the shortest path solves between the two nodes, i.e., consecutive base stop stations. We also need to create and store a look-up table for the paths corresponding to the shortest paths associated with this process-graph edge.

\section{Final SSSP}
\label{Section:Final}

Finally, a shortest path on the process-graph is calculated by running yet another Dijkstra solve but on this process-graph from source to target. The optimal path is the minimal cost aggregated over the consecutive shortest path runs among the stations implicitly. Hence, finding the shortest path on the process-graph is indeed the result of our optimization algorithm. Note that any edge whose weight is greater than the vehicle charging capacity is discarded, and not even inserted as an edge into the process graph as shown in Figure~\ref{Figure:usmatchpaths} with the red cross sign. The resulting shortest path is shown by a set of red line-segment in Figure~\ref{Figure:usmatchnetwork}. The path is aggregated with the paths of the shortest paths associated with the edge. Those aggregated paths are coming from the solves on the US network graph between each consecutive stations that we have cached and mapped to the edges of the process-graph as formulated by the Equation~\ref{Equation:finaldijkstra}. In the next Section~\ref{Section:Results}, we will demonstrate the SQL syntax of the graph solver endpoint call over several cross country trip examples to show the results of the optimal routing paths and stations.

\begin{align}
\mathscr{D}_{final} \mid_{start}^{end} = \min{ \left\{ \sum_{k=0}^{stops} {\left\| \mathscr{D}_{i^k,j^{k+1}} \right\|}  \right\}_{G} \forall i \neq j}
\label{Equation:finaldijkstra} 
\end{align}

\section{Results}
\label{Section:Results}

\begin{figure*}
\centering
    \framebox{\includegraphics[width=0.8\linewidth, keepaspectratio]{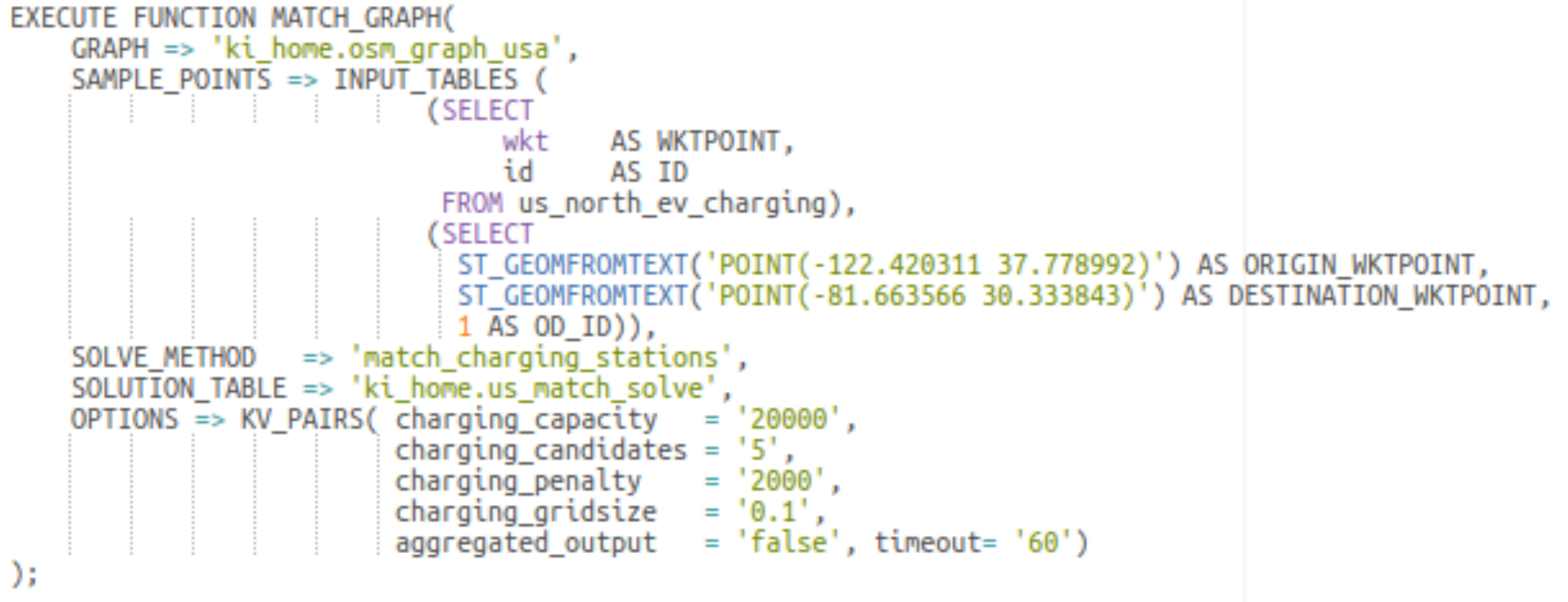}}
    \caption{SQL form of the graph solver endpoint Match-Graph request for the optimal path shown in Figure~\ref{Figure:usmatch10stations}. The input \textit{SAMPLE\_POINTS} component parameter is the available public stations database and the requested source and destination locations in longitude and latitude from San-Francisco to Jacksonville, FL, respectively. The unit of the charging options should be compatible with the unit of the graph weights, i.e., in this case, the weights are in seconds, hence, a charging capacity value of 20000 roughly translates to approximately 300 miles considering average highway speeds.}
    \label{Figure:usmatchgraph}
\end{figure*}

The optimization solver in this paper is an add-on solver to our existing Match-Graph graph endpoint (Restful API). The new solver is added to the list of other existing solver types, such as \textit{markov\_chain} for map matching, \textit{match\_supply\_demand} for multiple supply demand logistics, \textit{match\_loops} for Eulerian path detection etc., that is serviced by the same distributed Match-Graph endpoint as shown in Figure~\ref{Figure:usmatchgraph}.

\begin{figure*}
\centering
    \framebox{\includegraphics[width=\linewidth, keepaspectratio]{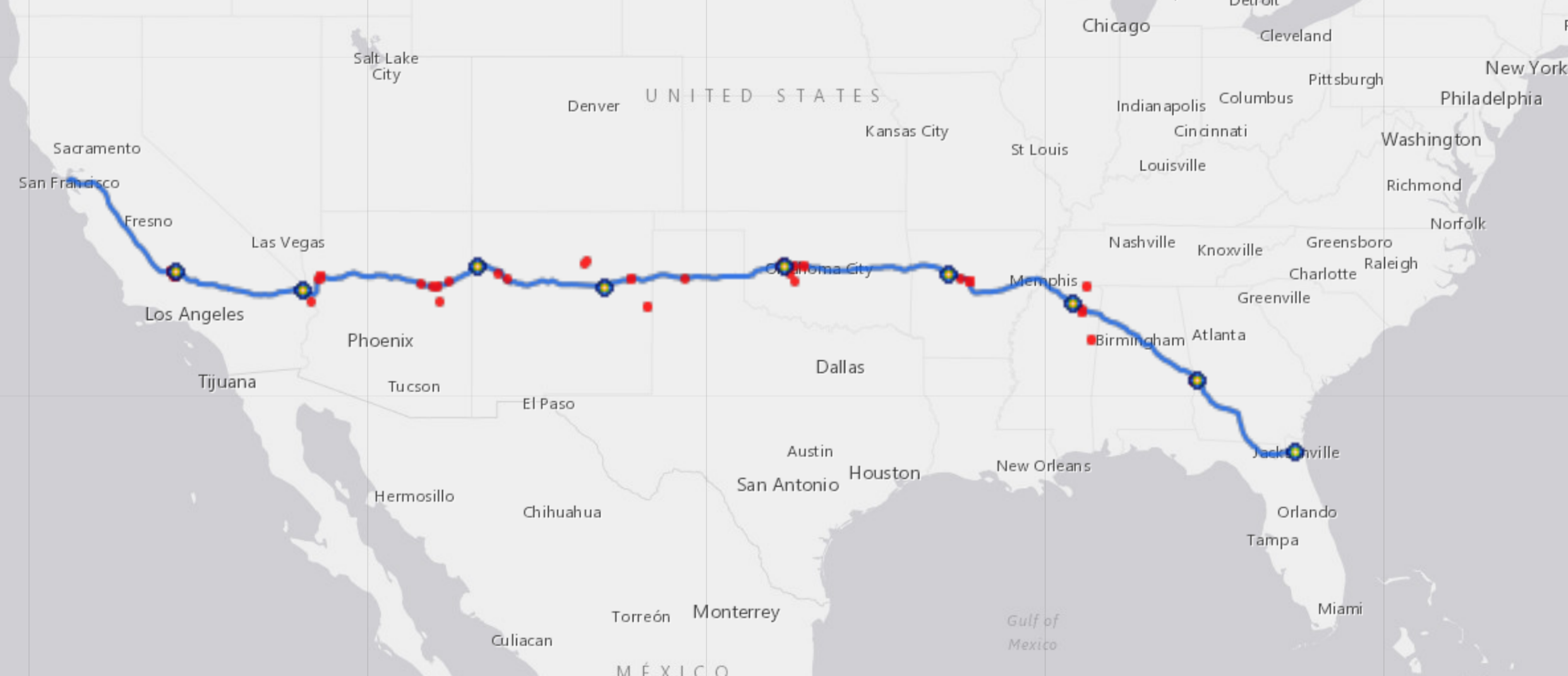}}
    \caption{ The result of optimal recharging route for a trip from San-Francisco to Jacksonville, FL using the SQL Match-Graph request of Figure~\ref{Figure:usmatchgraph} shown as a blue path. The red dots are the candidate stations found where recharging will be needed based on the capacity and the number of candidates passed as the input parameters to the solver. The solver then finds the optimal set of stations among these candidates at each recharging stop depicted as larger yellow circles so that the overall trip finishes in the shortest time including recharging and never exceeds the capacity.}
    \label{Figure:usmatch10stations}
\end{figure*}

The unit of the charging options should be compatible with the unit of the weights of the graph. The main solver parameters are the charging capacity of the vehicle and the full charging penalty. The database table for the EV public charging stations, including \textit{lon/lat} locations and the station \textit{ids} should also be provided with the appropriate Kinetica-Graph grammar as shown in Figure~\ref{Figure:usmatchgraph}. Another important aspect of our solver is that we also include exact charging time penalty into the optimization, i.e., if the cost of the aggregated Dijsktra, the edge weights in the process graph is not exactly requiring a full recharge at the station stop, we only add in the proportional amount of penalty that is required to top of the capacity as shown in the Equation~\ref{Equation:adjustedpenalty} where $cost_j^*$ is the adjusted edge weight of the process graph, and the $cost_j$ is coming form the sssp runs between the consecutive station pairs. 

\begin{align}
cost_j^* = cost_j +  charging\_penalty * cost_j/Q
\label{Equation:adjustedpenalty} 
\end{align}

\begin{figure}
\centering
    \framebox{\includegraphics[width=\linewidth, keepaspectratio]{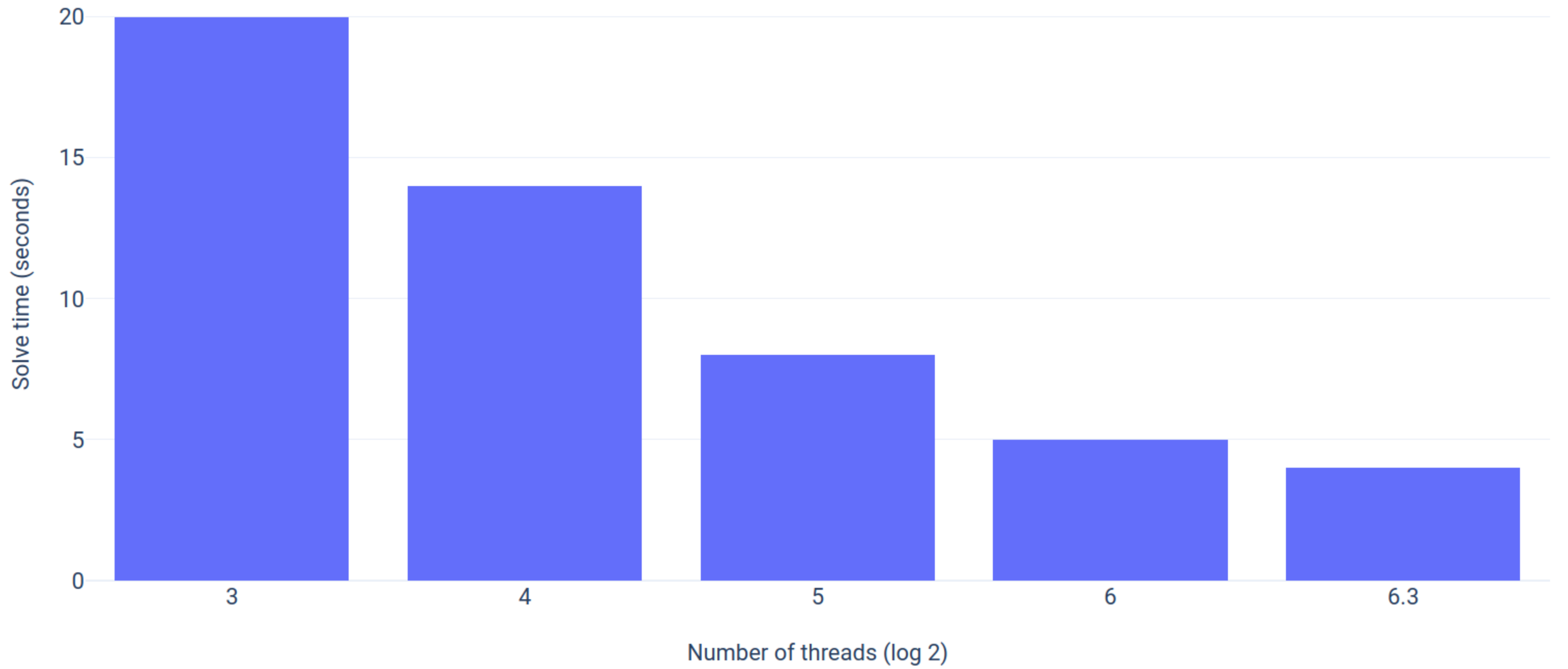}}
    \caption{The scalability chart of the solver; total solve time versus number of threads used (log 2 based). The scalable part of the solver is the sssp runs between each prospective stations pairs. Maximum number of cores used in this study is 80 (real cores), which corresponds to a total solve time of approximately 4-5 seconds including initial constant time A-star sssp, writing to output DB tables and the latency between the servers.}
    \label{Figure:usmatchscale}
\end{figure}

An example of a cross-country multi-stop (10 stations) ev-charging routing is demonstrated in Figure~\ref{Figure:usmatch10stations}, that has a slight deviation from the initial A-star sssp between the two end-points which is also a good self verification for the optimality of the routing. This deviation is more pronounced in another example shown in Figure~\ref{Figure:usmatchohio}. The trip planning in this instance requires 5 recharging stops and takes a bit longer than 18 hours as shown in the record of \textit{COST} column of the solution table in Figure~\ref{Figure:usmatchohiooutput}. The total solve time reduces with the number of threads used which can also be seen in Figure~\ref{Figure:usmatchscale}. The optimization for this 5 station stop case takes a bit more than 4 seconds on a 80 core machine. Hundreds of sssp runs are required for $n=m=5$ with 135 number of process graph edges calculated by Equation~\ref{Equation:processedges}. 

The optimization algorithm is a scalable solver due to the fact that these sssp Dijkstra runs among consecutive station stops  are  actually independent from each other. Moreover, revising the Dijkstra solver implementation aiming reduced memory and time SLA is really the major factor contributing to the effectiveness of the scalability within each thread's own sssp runs. These amortized runs are limited to operate within  a Dijkstra disk radius of only a percentage of the vehicle capacity, which also greatly reduces the number of edges involved in each solve to almost two orders of magnitude less than entire US graph of 160 million edges.

One possible future improvement on our Match-Graph optimization solver might be adding a conditional logic into the main algorithm in case searching around refueling locations would not be able to find any stations within the capacity limit. In that rare scenario, a possible mitigation technique could be to move the anticipated refueling location off the shortest path back and forth until one or more stations within the search radius can be spotted. Even though this mitigation technique is required to have the desired fail-safe status, in practice it almost never happens and predictably less so in the future as more ev-charging stations have continually been added in greater numbers even in the rural locations where the demand for EVs could only be assumed to increase exponentially in the near future.

\begin{figure}
\centering
    \framebox{\includegraphics[width=\linewidth, keepaspectratio]{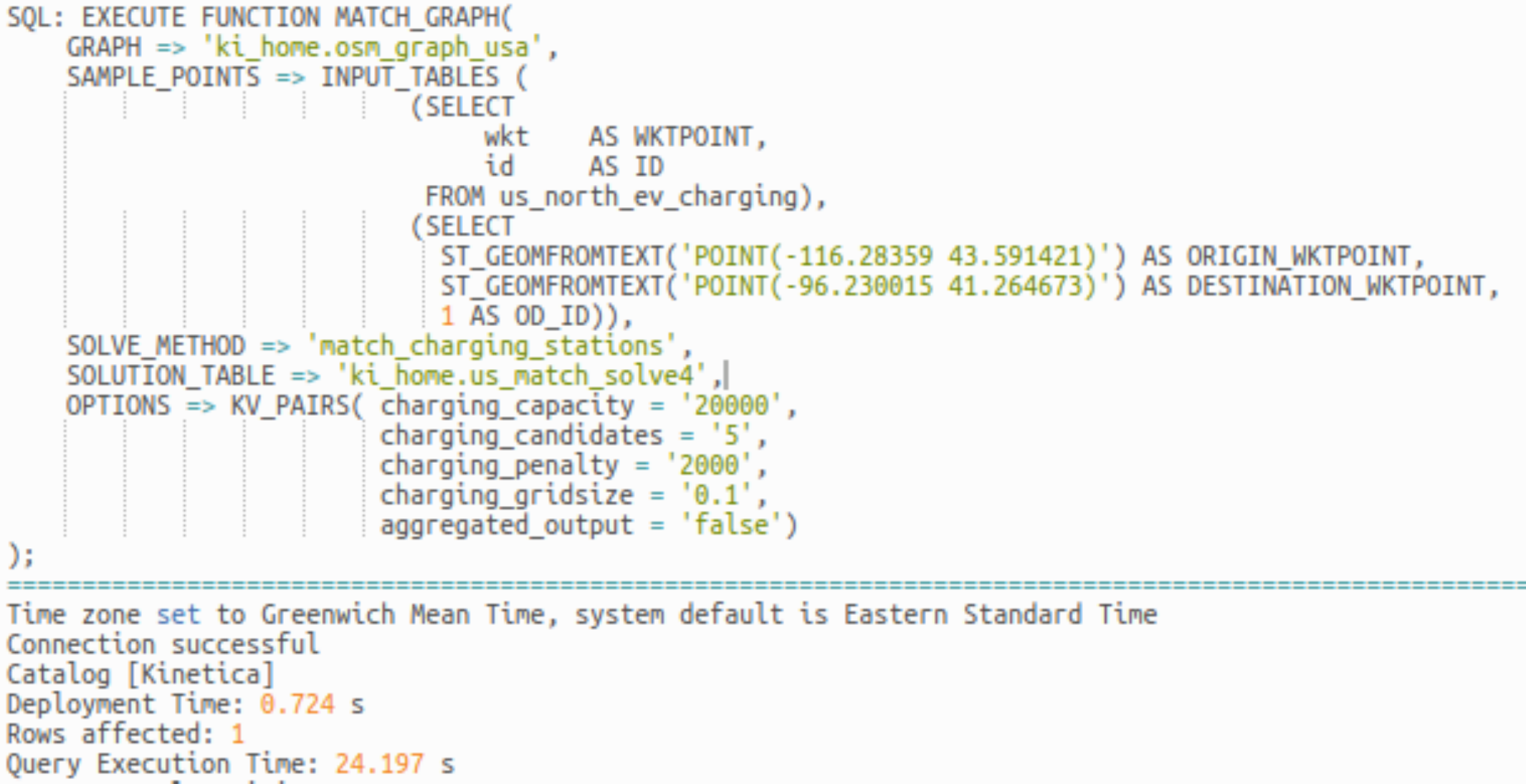}}
    \caption{SQL form of the graph solver endpoint Match-Graph request for the optimal path shown in Figure~\ref{Figure:usmatchohio}. Similar to the request in Figure~\ref{Figure:usmatchgraph}, the first parameter is the public EV charging stations database table, followed by the origin destination points depicted as geometry constants. The charging options for the capacity, and the maximum full charging time are specified to be around 300 miles, and 40 minutes, respectively. The charging candidates and gridsize parameters are actually internal options.}
    \label{Figure:usmatchohiosolve}
\end{figure}

\begin{figure}
\centering
    \framebox{\includegraphics[width=\linewidth, keepaspectratio]{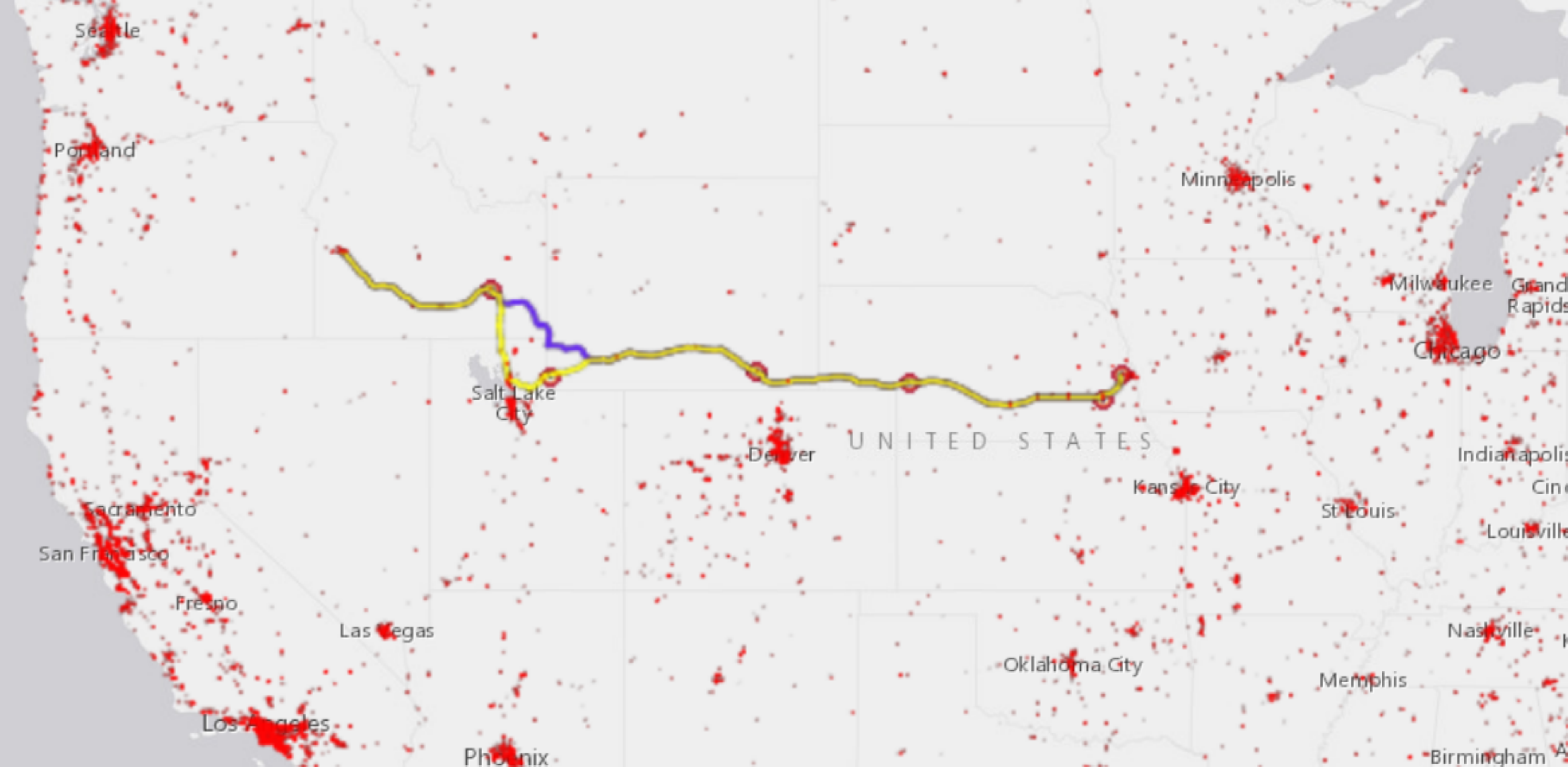}}
    \caption{The result of optimal recharging route for a trip that requires approximately 19 hours of driving. The red dots are all the available stations, and the red hallow circles are the optimal stations found by the solver for recharging. The blue path is the shortest path from source to target for comparison. There are a total of five stations computed along the way as depicted in Figure~\ref{Figure:usmatchohiosolve}.}
    \label{Figure:usmatchohio}
\end{figure}

\begin{figure}
\centering
    \framebox{\includegraphics[width=\linewidth, keepaspectratio]{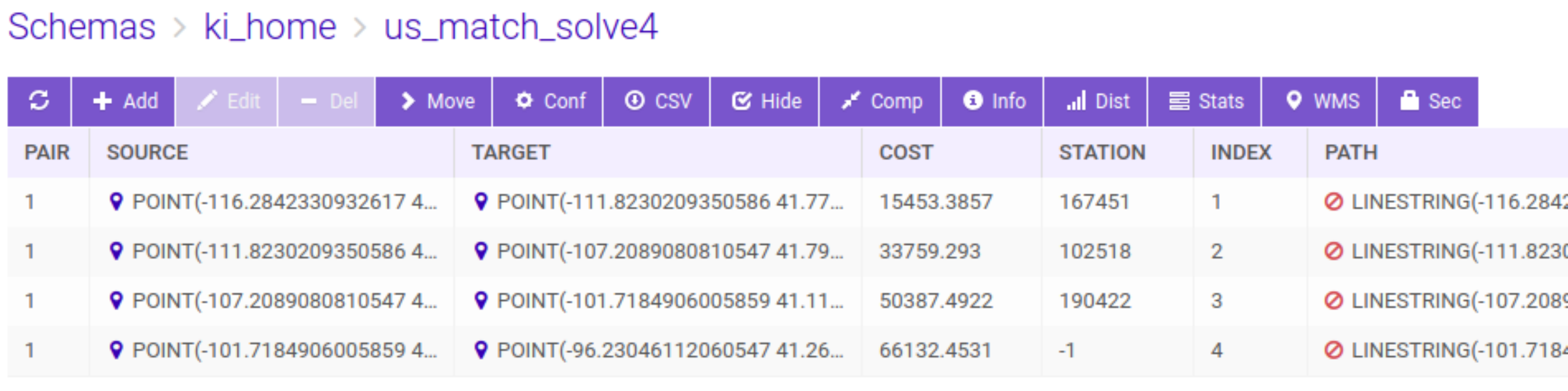}}
    \caption{ The resulting solution table of the Match-Graph request depicted in Figure~\ref{Figure:usmatchohiosolve}. The request is made with the \textit{aggregate\_output} option of \textit{false}, and hence the format includes each leg of the trip at a different record with the corresponding trip path index. The station ids on column \textit{STATION} matches with the input \textit{id} column the public EV stations database table.}
    \label{Figure:usmatchohiooutput}
\end{figure}

\section*{Acknowledgement}

The authors would like to thank the technical contributions of the entire Kinetica Engineering team, and more specifically, Rydel Pereira for his help on retrieving the EV public charging stations data and finally our CEO Nima Negahban for his strong support of Kinetica-Graph since its inception.

\section*{Notes on Contributors}
\small{
\noindent \textbf{Bilge Kaan Karamete} is the lead technologist for the Geospatial, Graph and Visualization efforts at Kinetica. His research interests include computational algorithm development, unstructured mesh generation, parallel graph solvers and computational geometry. He holds a PhD in Engineering Sciences from the Middle East Technical University, Ankara Turkey, and post doctorate in Computational Sciences from Rensselaer Polytechnic Institute, Troy New York.

\noindent \textbf{Eli Glaser} is SVP of Engineering at Kinetica. He leads the development teams concentrating in data analytics, query capability and performance. Eli holds a Master's in Electrical Engineering from The Johns Hopkins University, Baltimore Maryland.
}


\section{Software avaliability}

Kinetica's Developer Edition is freely available here https://www.kinetica.com/try/.

\section*{References}

\bibliography{ev}

\end{document}